\documentclass[nofootinbib]{revtex4}
\usepackage{graphicx}
\usepackage{dcolumn}
\usepackage{amsmath,amssymb,epsfig}
\usepackage{paralist}
\usepackage{comment}
\usepackage{graphicx}
\usepackage{wrapfig}
\usepackage{multirow}
\usepackage{color,soul}
\usepackage{suffix}
\usepackage{mathtools}

\allowdisplaybreaks

\renewcommand{\vec}[1]{\boldsymbol{\mathrm{#1}}}

\begin{document}

\title{Image formation process with the solar gravitational lens}

\author{Slava G. Turyshev$^{1}$, Viktor T. Toth$^2$}

\affiliation{\vskip 3pt
$^1$Jet Propulsion Laboratory, California Institute of Technology,\\
4800 Oak Grove Drive, Pasadena, CA 91109-0899, USA}

\affiliation{\vskip 3pt
$^2$Ottawa, Ontario K1N 9H5, Canada}

\date{\today}

\begin{abstract}

We study image formation with the solar gravitational lens (SGL). We consider a point source that is positioned at a large but finite distance from the Sun. We  assume that an optical telescope is positioned in the image plane, in the focal region of the SGL. We model the telescope as a convex lens and evaluate the intensity distribution produced by the electromagnetic field that forms the image in the focal plane of the convex lens. We first investigate the case when the telescope is located on the optical axis of the SGL or in its immediate vicinity. This is the region of strong interference where the SGL forms an image of a distant source, which is our primary interest. We derive analytic expressions that describe the progression of the image from an Einstein ring corresponding to an on-axis telescope position, to the case of two bright spots when the telescope is positioned some distance away from the optical axis. At greater distances from the optical axis, in the region of weak interference and that of geometric optics, we recover expressions that are familiar from models of gravitational microlensing, but developed here using a wave-optical treatment. We discuss applications of the results for imaging and spectroscopy of exoplanets with the SGL.

\end{abstract}


\maketitle

\section{Introduction}
\label{sec:aintro}

According to Einstein's general theory of relativity, as light travels in the vicinity of the Sun, light ray trajectories are bent towards the Sun by an angle of $\theta=2r_g/b=1.75\, (R_\odot/b)$ arcseconds, where $r_g$ is the Schwarzschild radius of the Sun, $b$ is the trajectory's impact parameter and $R_\odot$ is the solar radius. In this context, the Sun acts a lens by focusing light at heliocentric ranges beyond $b^2/(2r_g)=547.6\,(b/R_\odot)^2$ AU, by amplifying its brightness by a factor of $4\pi^2 r_g/\lambda=2.11\times 10^{11}\,(1\,\mu{\rm m}/\lambda)$, where $\lambda$ is the observable wavelength, and also by naturally providing an angular resolution of $\lambda/(2R_\odot)\sim 0.2 \,(\lambda/1\,\mu{\rm m})$ nanoarcseconds  \cite{Turyshev:2017,Turyshev-Toth:2017}. This behavior of the solar gravity field is known as the solar gravitational lens (SGL)  \cite{Turyshev:2017,Turyshev-Toth:2017,Turyshev-Toth:2019}. Although the focal region of the SGL begins at large heliocentric distances, successful deep space missions such as Voyager 1/2 and Pioneer 10/11 demonstrated that the capability exists to build spacecraft that can travel to the SGL focal region, operate there successfully, while maintaining reliable communication with the Earth. This opens up the possibility of using the SGL to build an astronomical facility with tremendous light gathering power and angular resolution \cite{Turyshev-etal:2018}.

The most conventional concept of exploiting the SGL envisions an optical telescope in the SGL's image forming region, looking in the direction of the Sun and observing the Einstein ring formed by light from a distant source around the Sun. As light from the Sun itself will likely dominate any faint light from a distant source, the optical telescope must have sufficient angular resolution to resolve the solar disk, to make it possible to block sunlight using a coronagraph. This necessitates the use of a telescope with a meter-class  or larger aperture.

Conceptually, then, the study of the SGL can be broken down into several discrete phases, beginning with the study of plane waves (i.e., light from a point source at infinity) deflected by a spherical, transparent Sun \cite{Turyshev-Toth:2017}. Building on this foundation, we can introduce the opaque solar disk and its shadow region \cite{Turyshev-Toth:2018,Turyshev-Toth:2018-grav-shadow}, we can study the effects on light by the solar corona \cite{Turyshev-Toth:2018-plasma,Turyshev-Toth:2019}, and eventually, we can extend our efforts to study light from extended sources at a finite distance and the resulting image formed by the SGL \cite{Turyshev-Toth:2019-extend}. That work led to  \cite{Turyshev-Toth:2019-blur} where we obtained expressions to characterize the power of the signal that is received by a telescope from a distant object, such as an exoplanet.

The next step is to consider the intensity distribution pattern of this signal that appears in the focal plane of the optical telescope, as a function of the telescope's displacement from the optical axis of the SGL. Therefore, our goal in the present paper is to investigate what such an imaging telescope ``sees'' both in the immediate vicinity of the focal line (the imaginary line connecting the distant source to the center of the Sun and extending towards the focal region on the other side of the Sun) and also far from the focal line, in the geometric optics region. The former is of great interest for direct high-resolution imaging and spectroscopy of faint sources with the SGL; the latter establishes a direct connection between our work and studies of gravitational microlensing. This is done to demonstrate the power of the wave optical treatment in describing the scattering of light on a gravity field. Ultimately, our objective is to derive expressions that can be used to model anticipated signals from realistic sources, focused and amplified by the SGL.  This is needed for both to evaluate the potential science return from a deep-space mission to the focal region of the SGL and also to develop a set of requirements needed to design such a mission \cite{Turyshev-etal:2018}.

Our paper is organized as follows:
In Section~\ref{sec:image-sens}, we provide the general approach on how to evaluate the signal that is observed in the focal plane of a telescope and how to evaluate the corresponding intensity distribution pattern in that focal plane.
Section~ \ref{sec:image-point} introduces the SGL and the solution for the  electromagnetic (EM) field in the image plane. We discuss modeling  the signal from a point source as it is received in the image plane of an optical telescope. We consider cases with both small and large departures from the optical axis, while staying within the strong interference region of the SGL.
In Section~\ref{sec:go-wint} we consider imaging in the geometric optics and weak interference regions.
In Section \ref{sec:disc} we discuss results and avenues for the next phase of our investigation of the SGL.

\section{Image formation by an optical telescope}
\label{sec:image-sens}

\begin{figure}[t]
\includegraphics[scale=0.27]{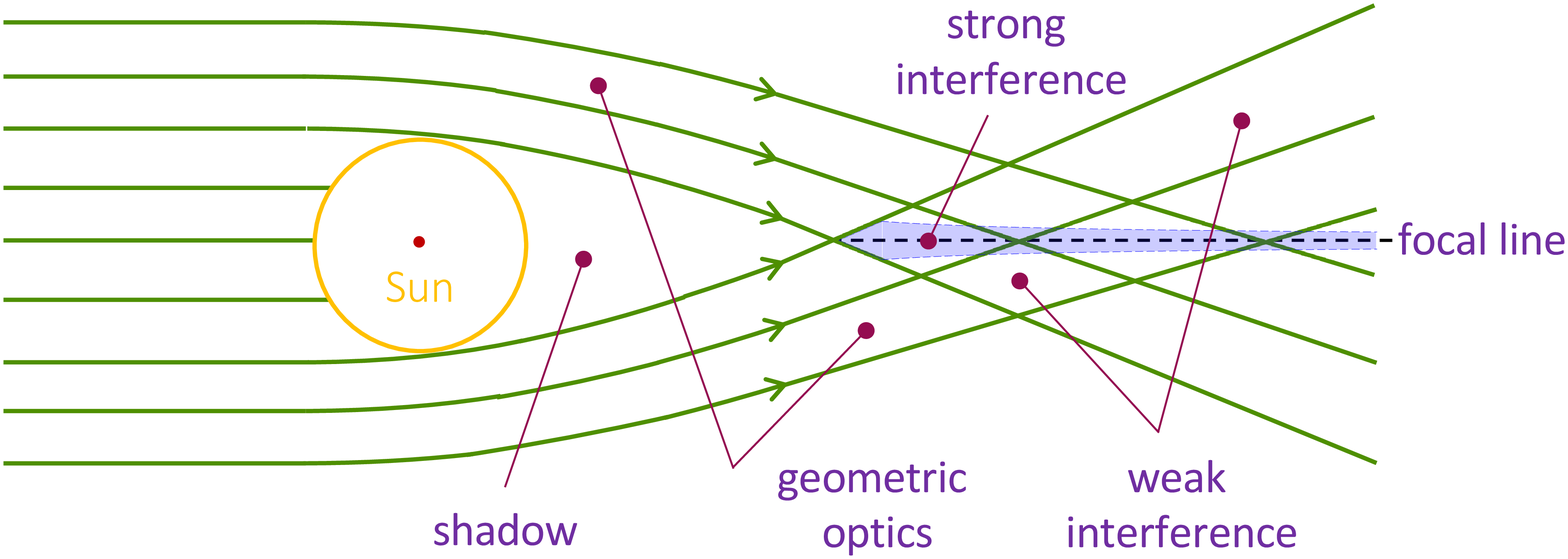}
\caption{\label{fig:regions}The different optical regions of the SGL
(adapted from \cite{Turyshev-Toth:2019-extend}).
}
\end{figure}

\begin{figure}[h]
\includegraphics[scale=0.65]{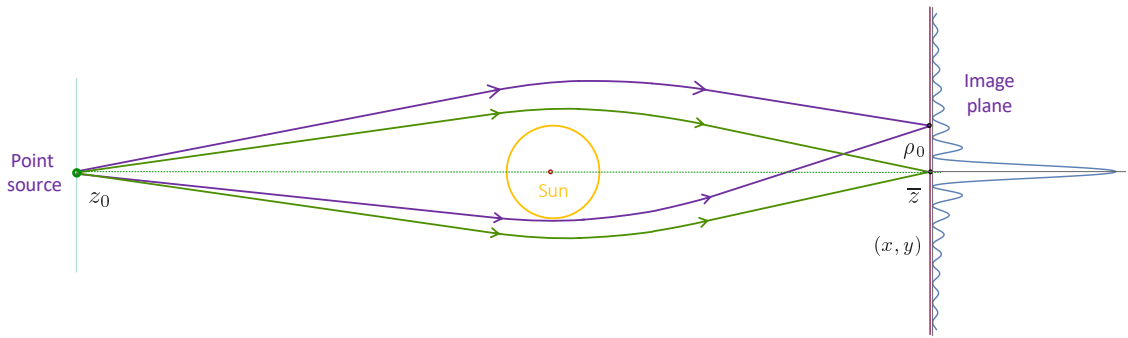}
\caption{\label{fig:imaging-geom}Imaging a point source with the SGL. The source is positioned on the optical axis at the distance $z_0$ from the Sun. The image plane is at the heliocentric distance $\overline z$. Rays with different optical paths produce a diffraction pattern in the image plane that is observed by an imaging telescope.}
\end{figure}

To describe the imaging process with the SGL, we position an imaging telescope in the strong interference region formed behind the Sun in the immediate vicinity of the SGL's optical axis. In the case of a point source, the optical axis is an imaginary line that connects the point source and the center of the Sun and extends behind the Sun into strong interference region, see Fig.~\ref{fig:regions}. This is the region where the SGL forms an image of a distant source. We take the source to be at the distance of $z_0$ from the Sun and position the telescope at heliocentric distance $z$, see Fig.~\ref{fig:imaging-geom}.

The relevant geometry is described by several parameters, such as ${\vec x}_0$ being the current position of an optical telescope in the SGL's image plane, ${\vec x}$, being any point on the same plane, and ${\vec x}_i$, being a point on the focal plane of the optical telescope. These positions are given as
{}
\begin{eqnarray}
\{{\vec x}_0\}&\equiv& (x_0,y_0)=\rho_0\big(\cos\phi_0,\sin\phi_0\big)=\rho_0{\vec n}_0, \label{eq:coord'}\\
\{{\vec x}\}&\equiv& (x,y)=\rho\big(\cos\phi,\sin\phi\big)=\rho\,{\vec n},
\label{eq:x-im} \\
 \{{\vec x}_i\}&\equiv& (x_i,y_i)=\rho_i\big(\cos\phi_i,\sin\phi_i\big)=\rho_i{\vec n}_i.
  \label{eq:coord}
\end{eqnarray}

\begin{figure}[t]
\includegraphics[scale=0.72]{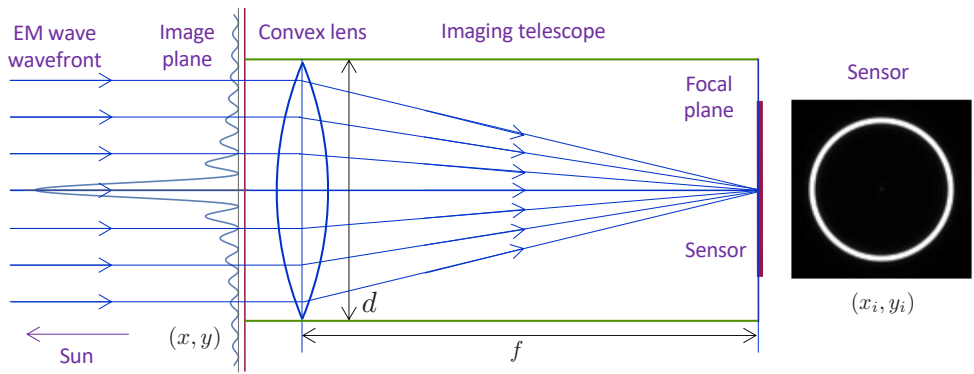}
\caption{\label{fig:imaging-sensor}Imaging a point source with the SGL as seen by a telescope. The telescope is represented by a convex lens with aperture $d$ and a focal length $f$. The telescope optics projects an image of the Einstein ring around the Sun in its focal plane. Positions in the SGL image plane, $(x,y)$, and the telescope's focal plane, $(x_i,y_i)$, are also shown.}
\end{figure}

To produce images with the SGL, we represent an imaging telescope by a  convex lens with focal distance $f$ and position the telescope in the interference region (see Figs.~\ref{fig:regions}--\ref{fig:imaging-sensor}). Following \cite{Born-Wolf:1999}, the EM field at a particular location ${\vec x}_i=(x_i,y_i)$ in the focal plane of the lens is given as
{}
\begin{eqnarray}
    \left( \begin{aligned}
{E}_\rho& \\
{H}_\rho& \\
  \end{aligned} \right)
   =    \left( \begin{aligned}
{H}_\phi& \\
-{E}_\phi& \\
  \end{aligned} \right)
   &=&
   E_0 {\cal A}({\vec x}_i,{\vec x}_0)
 \left( \begin{aligned}
 \cos\phi& \\
 \sin\phi& \\
  \end{aligned} \right)    e^{i\Omega (r,t)},
  \label{eq:DB-sol-rho2}
\end{eqnarray}
where $\Omega (r,t)$ is the time-dependent phase of a plane wave.

The quantity ${\cal A}({\vec x}_i,{\vec x}_0)$ in (\ref{eq:DB-sol-rho2}) is the complex amplitude of the EM wave as it observed on the focal plane of the optical telescope for a particular telescope position of $\vec{x}_0$. If the amplitude at the entrance of the telescope, ${\cal A}({\vec x},{\vec x}_0)$, is known for any location $\vec{x}$ on the image plane (in the case of imaging with the SGL, this amplitude is well-known, e.g., see \cite{Turyshev:2017,Turyshev-Toth:2017,Turyshev-Toth:2019-extend}), then the wave's amplitude at the focal plane of the telescope is determined by the Fresnel--Kirchhoff diffraction formula \cite{Born-Wolf:1999}:
{}
\begin{eqnarray}
{\cal A}({\vec x}_i,{\vec x}_0)=\frac{i}{\lambda}\iint \displaylimits_{|{\vec x}|^2\leq (d/2)^2} \hskip -7pt  {\cal A}({\vec x},{\vec x}_0)e^{-i\frac{k}{2f}|{\vec x}|^2}\frac{e^{iks}}{s}d^2{\vec x}.
  \label{eq:amp-w-f0}
\end{eqnarray}

The function $e^{-i\frac{k}{2f}|{\vec x}|^2}=e^{-i\frac{k}{2f}(x^2+y^2)}$ represents the action of the convex lens that transforms incident plane waves into spherical waves focused at the focal point. Assuming that the focal length is sufficiently large compared to the telescope aperture, we may approximate the optical path $s$ as $s=\sqrt{(x-x_i)^2+(y-y_i)^2+f^2}\sim f+\big((x-x_i)^2+(y-y_i)^2\big)/2f$. This allows us to present (\ref{eq:amp-w-f0}) as
{}
\begin{eqnarray}
{\cal A}({\vec x}_i,{\vec x}_0)&=&
- \frac{e^{ikf(1+{{\vec x}_i^2}/{2f^2})}}{i\lambda f}\iint\displaylimits_{|{\vec x}|^2\leq (\frac{1}{2}d)^2} d^2{\vec x}\,
{\cal A}({\vec x},{\vec x}_0) e^{-i\frac{k}{f}({\vec x}\cdot{\vec x}_i)}.
  \label{eq:amp-w-f}
\end{eqnarray}
Therefore, the presence of a convex lens  is equivalent to a Fourier transform of the complex wave amplitude \cite{Born-Wolf:1999}.

Using these results, we can compute the Poynting vector for the EM field emitted  by a point source and received at ${\vec x}_i$ in the telescope image plane. Given the form of the EM field, (\ref{eq:DB-sol-rho2}), the Poynting vector will have only one nonzero component, $S_z$. Using the overbar to denote time-averaging, we compute $S_z$ from (\ref{eq:DB-sol-rho2}) as
 {}
\begin{eqnarray}
S_z({\vec x}_i,{\vec x}_0)=\frac{c}{4\pi}\overline{[{\rm Re}{\vec E}\times {\rm Re}{\vec H}]}=\frac{cE_0^2}{4\pi} \overline{\big({\rm Re} [{\cal A}({\vec x}_i,{\vec x}_0)e^{i\Omega(t)}]\big)^2}.
  \label{eq:Pv}
\end{eqnarray}

As a result, once the amplitude of the EM field in the telescope image plane, ${\cal A}({\vec x}_i,{\vec x}_0)$, is known, using (\ref{eq:Pv}) we can compute the energy deposited in the image plane and evaluate the corresponding intensity distribution on the focal plane of the telescope that constitutes the observed image.

\section{Modeling the EM signal in the focal plane}
\label{sec:image-point}

\subsection{EM field in the interference region of the SGL}
\label{sec:EM-field}

In \cite{Turyshev-Toth:2019-extend}, we considered light from an extended source at a finite distance, $r_0$ from the Sun. We parameterize the problem using a spherical coordinate system $(r,\theta,\phi)$ that is aligned with a preferred axis: a line connecting a preselected (e.g., central) point in the source to the center of the Sun. We also use a cylindrical coordinate system $(\rho,z,\phi)$, with the $z$-axis corresponding to the preferred axis. Also, we characterize points in the SGL image plane according to (\ref{eq:coord'})--(\ref{eq:coord}). Here we are interested in the details of the image formation process and, thus, consider imaging of a point source.

To describe the imaging process, we consider the EM field in the strong interference region  just in front of the telescope. For that, we consider light, that is to say, a high-frequency EM wave (i.e., neglecting terms $\propto(kr)^{-1}$ where $k$ is the wavenumber) and for $r\gg r_g$ (where $r_g=2GM_\odot/c^2$ is the Sun's Schwarzschild radius) and derive the components of the EM field near the optical axis. Following \cite{Turyshev-Toth:2019-extend}, we have that for a point source located on the optical axis at the distance of $z_0$ from the Sun, up to terms of ${\cal O}(\rho^2/z^2, \sqrt{2r_g  \overline z}/z_0)$, the $z$-component of the EM field which moves in the $z$-direction (Fig.~\ref{fig:regions}) behaves as $({E}_z, {H}_z)={\cal O}({\rho}/{z},\sqrt{2r_g\overline z}/z_0)$, while the other components are given as
{}
\begin{eqnarray}
    \left( \begin{aligned}
{E}_\rho& \\
{H}_\rho& \\
  \end{aligned} \right) =    \left( \begin{aligned}
{H}_\phi& \\
-{E}_\phi& \\
  \end{aligned} \right)&=&
  E_0 {\cal A}({\vec x},{\vec x}_0)
 \left( \begin{aligned}
 \cos\phi& \\
 \sin\phi& \\
  \end{aligned} \right) e^{i\Omega (t)},\qquad
  \Omega (t)=k\big(r+r_0+r_g\ln 2k(r+r_0)\big)-\omega t,
  \label{eq:DB-sol-rho}
\end{eqnarray}
where the amplitude of the EM wave is given as \cite{Turyshev-Toth:2017,Turyshev-Toth:2019-extend}
{}
\begin{eqnarray}
{\cal A}({\vec x},{\vec x}_0)&=&
\sqrt{\mu_0} J_0\Big(k
\sqrt{\frac{2r_g}{\overline z}} |{\vec x}+{\vec x}_0|\Big), \qquad
\mu_0= 2\pi k r_g.
  \label{eq:mu}
\end{eqnarray}
Note that these expressions are valid for forward scattering when $\theta+b/z_0\ll \sqrt{2r_g/r}$, or when the deviation from the optical axis is small, $0\leq \rho\lesssim r_g$.

Substituting the complex amplitude ${\cal A}({\vec x},{\vec x}_0)$ from (\ref{eq:mu}) into expression (\ref{eq:amp-w-f}), we see that the amplitude of the EM field on the focal plane of an imaging telescope takes the following form:
{}
\begin{eqnarray}
{\cal A}({\vec x}_i,{\vec x}_0)&=&-\sqrt{\mu_0}\frac{e^{ikf(1+{{\vec x}_i^2}/{2f^2})}}{i\lambda f}
\hskip -8pt
 \iint\displaylimits_{|{\vec x}|^2\leq (d/2)^2}\hskip -8pt
  d^2{\vec x}
  J_0\big(\alpha |{\vec x}+{\vec x}_0|\big) e^{-i\eta_i({\vec x}\cdot{\vec n}_i)},
  \label{eq:amp-w-d*}
\end{eqnarray}
where we introduced the following convenient notations for the two relevant  spatial frequencies:
{}
\begin{eqnarray}
\alpha=k \sqrt{\frac{2r_g}{\overline z}}, \qquad \eta_i=k\frac{\rho_i}{f},
  \label{eq:alpha-mu}
\end{eqnarray}
where $\alpha$ describes the spatial frequency corresponding to the fixed position of the Einstein ring and $\eta_i$ is the variable spatial frequency for a particular location on the imaging sensor. As we see below, the interplay between these two frequencies governs the image formation process in the strong interference region of the SGL.

First, we observe that for optical frequencies, the spatial frequency $\alpha$ introduced by (\ref{eq:alpha-mu}) is rather high, behaving as $\alpha=48.98 \,(1\,\mu{\rm m}/\lambda)({650\,{\rm AU}/{\overline z}})^{\frac{1}{2}} ~{\rm m}^{-1}$. We also note that the size of the combination of $(\alpha\rho_0)$ is what determines the behavior of the Bessel function in (\ref{eq:amp-w-d*}). This function's behavior offers a natural approach to define three regions exhibiting different optical properties:
\begin{enumerate}[1)]
\item Very small deviations, characterized by $\alpha \rho_0\ll1$, representing
displacements from the optical axis  where $\rho_0$ is in the range $0\leq \rho_0\ll 1/\alpha=2.04 \times 10^{-2}\, \,(\lambda/1\,\mu{\rm m}){({\overline z}/650\,{\rm AU}})^{\frac{1}{2}} \, {\rm m}\approx 2\,{\rm cm}$. This is the case when the permitted displacements are within the central peak of $J_0$, which is well within the nominal telescope aperture of $d=1\,{\rm m}$;
\item moderate deviations, $\alpha \rho_0\approx 1$, representing displacements in the range of $1/\alpha\leq \rho_0\approx 10/\alpha\approx 20\,({\overline z}/650\,{\rm AU})^{\frac{1}{2}} \,{\rm cm}$, also still less than $d/2$; and finally
\item large deviations, $\alpha \rho_0\gg1$, described as $\rho_0> 10/\alpha\approx 20\,({\overline z}/650\,{\rm AU})^{\frac{1}{2}} \,{\rm cm}$.
\end{enumerate}

The second case is interesting, but difficult to study analytically. If needed, this case can be studied using numerical evaluation of the integral in (\ref{eq:amp-w-d*}). We also note that this case is of limited practical importance for imaging with the SGL, where we the use of telescopes with meter-class apertures \cite{Turyshev-etal:2018}.

Fortunately, given the fact that the spatial frequency $\alpha$ is rather large, for a very small displacement $\rho_0$ from the optical axis, there is a rapid transition between the regimes of the first and third case. Therefore, without a significant loss of generality we can restrict our study to these two cases: that is,
\begin{inparaenum}[i)]
\item the case when the telescope is positioned at a very small distance with respect to the optical axis or $\rho_0\ll d$ and
\item the case when the displacement of the telescope is large or $\rho_0\geq d$.
\end{inparaenum}
To be more specific, the relevant regimes are given by the relationships $\alpha r_0\ll 1$ and $\alpha \rho_0\gg 1$. Based on the analysis of the point-spread function (PSF) in \cite{Turyshev-Toth:2017},  to satisfy the condition $\alpha \rho_0\ll1$, the displacement from the optical axis must be very small, $0<\rho_0\ll 2\,({\overline z}/650\,{\rm AU})^{\frac{1}{2}}$~cm, all within the central peak of the PSF. Starting from  $\rho_0\gtrsim 0.5\,({\overline z}/650\,{\rm AU})^{\frac{1}{2}} $~m, we enter the regime of $\alpha \rho_0 \gg1$.

In the next two subsections, we develop approximate solutions to (\ref{eq:amp-w-d*}) for these two regimes with their distinct behavior.

\subsection{Small displacements from the optical axis}
\label{sec:small-disp}

We first consider the situation of small telescope displacements from the optical axis,  $|{\vec x}_0|\ll d$, also preserving the inequality, $\alpha \rho_0\ll1$. To evaluate the integral (\ref{eq:amp-w-d*}), utilizing the fact that $\rho_0\ll|{\vec x}|$,  we expand $ |{\vec x}+{\vec x}_0|$ to first order in ${\vec x}_0$ in the argument of the Bessel function:
 {}
\begin{eqnarray}
 |{\vec x}+{\vec x}_0|=\rho+{\cal O}(\rho_0).
  \label{eq:mod0}
\end{eqnarray}
This allows us to evaluate the resulting integral:
{}
\begin{eqnarray}
\int_0^{d/2}\hskip -12pt \rho d\rho  \int_0^{2\pi}  \hskip -10pt d\phi\,
  J_0(\alpha \rho)  e^{-i\eta_i \rho\cos(\phi-\phi_i)}=
\pi\Big(\frac{d}{2}\Big)^2 \bigg(\frac{2}{(\alpha^2-\eta_i^2){\textstyle\frac{1}{2}}d}
\Big( \alpha J_0(\eta_i {\textstyle\frac{1}{2}}d) J_1(\alpha {\textstyle\frac{1}{2}}d)-\eta_i J_0(\alpha {\textstyle\frac{1}{2}}d) J_1(\eta_i {\textstyle\frac{1}{2}}d)\Big)+{\cal O}\big(\alpha\rho_0\big)\bigg).~
  \label{eq:amp-B-res}
\end{eqnarray}

\begin{figure}
\includegraphics[width=0.3\linewidth]{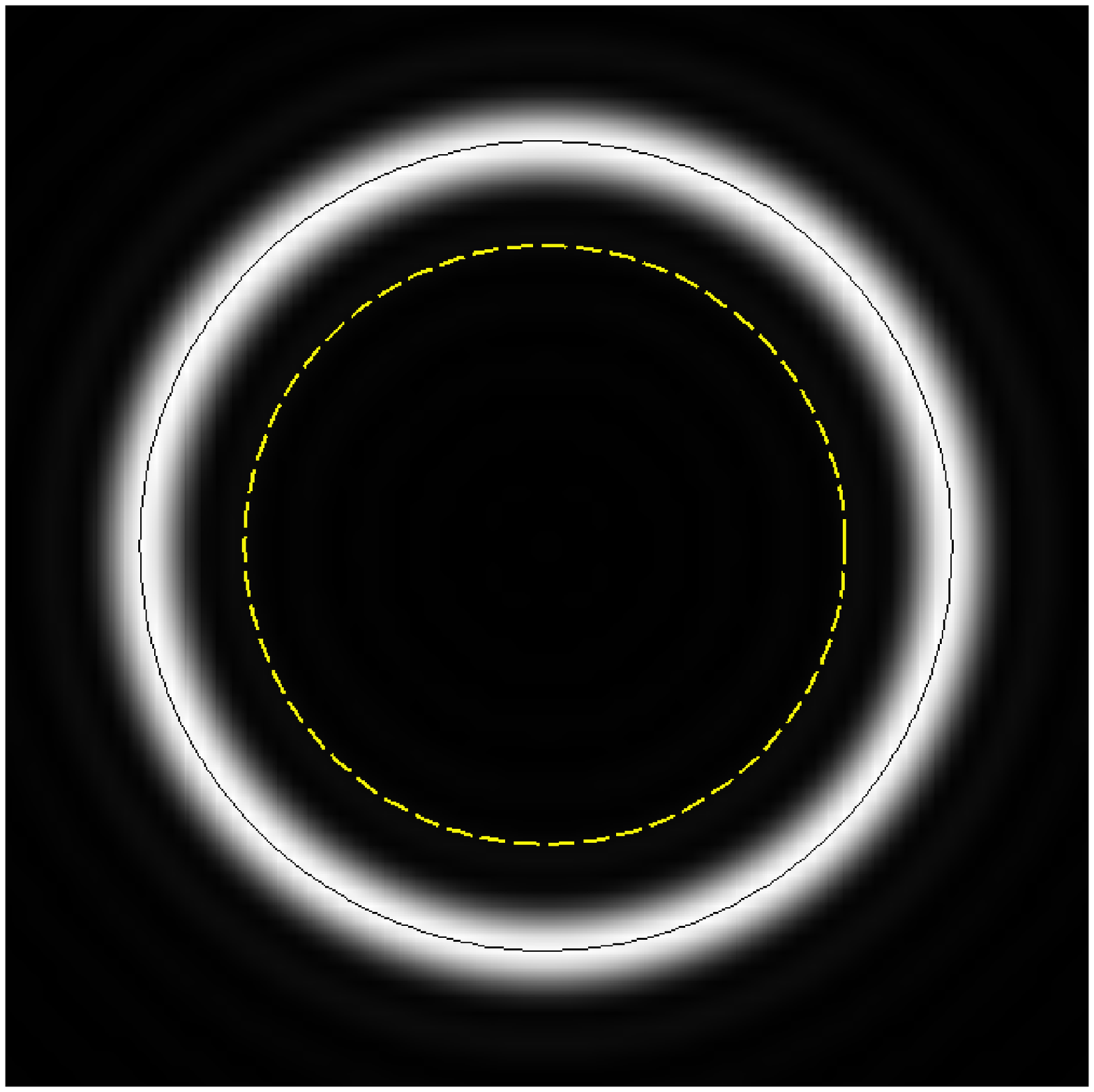}~\includegraphics[width=0.3\linewidth]{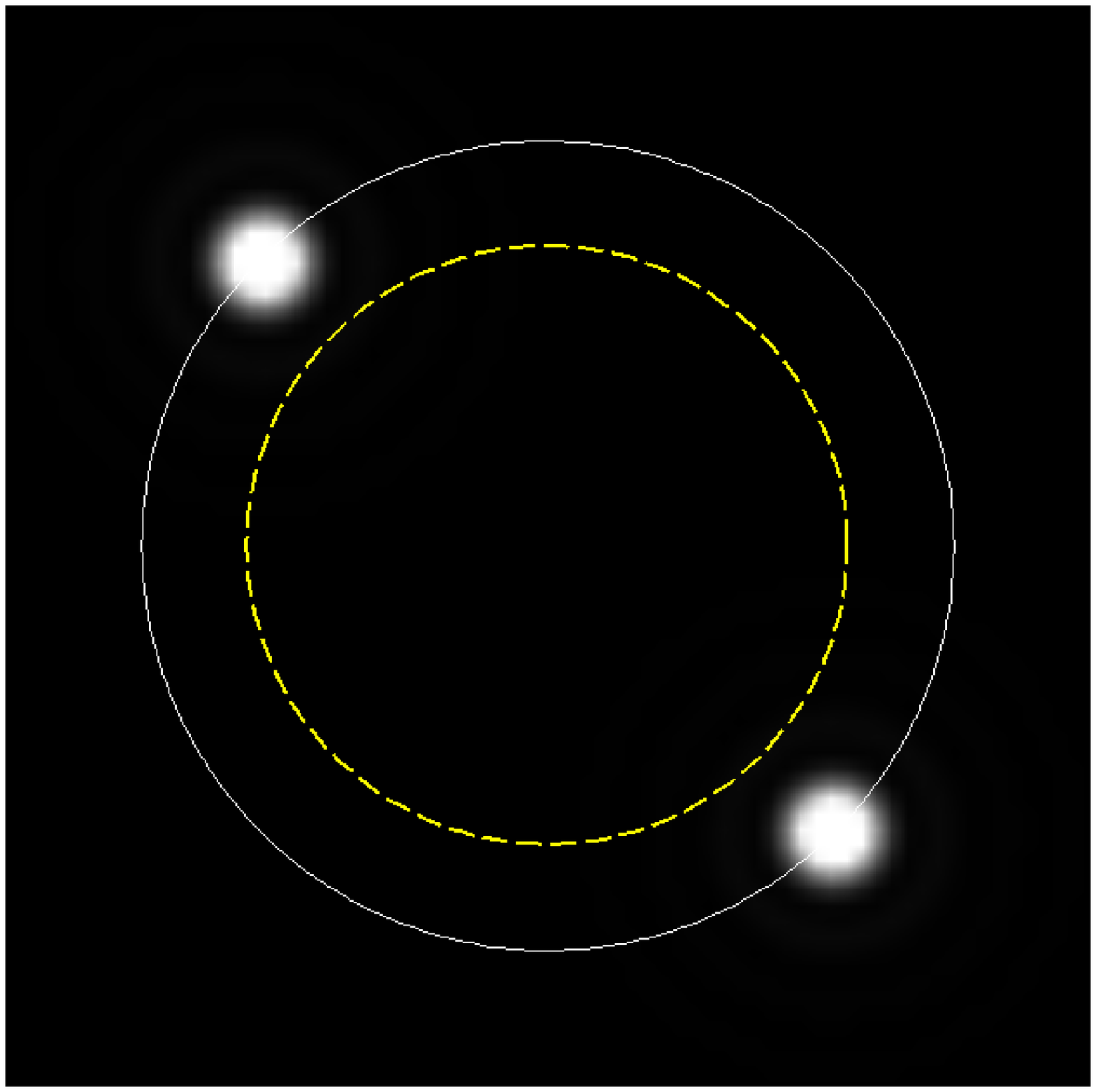}~\includegraphics[width=0.3\linewidth]{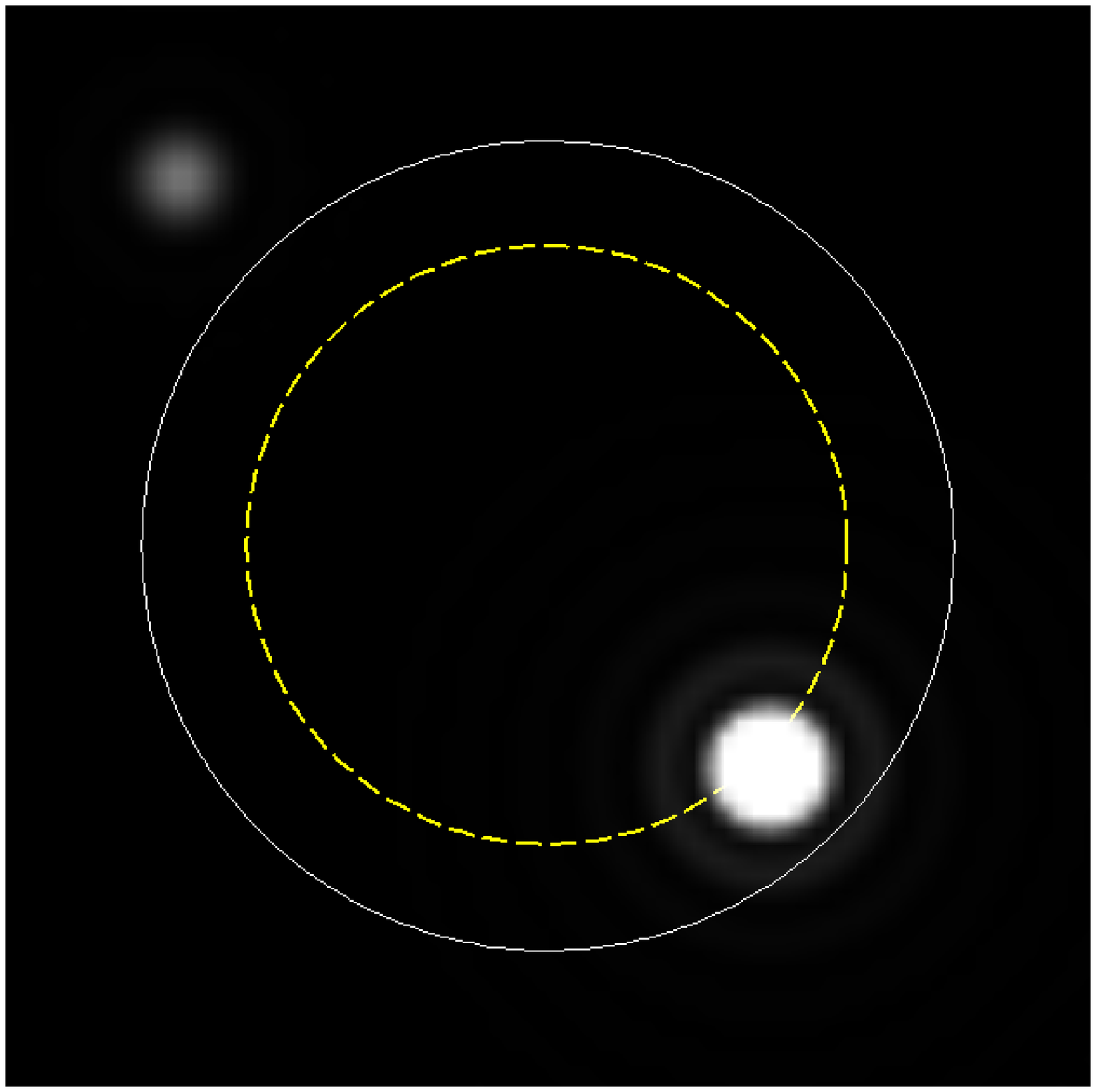}
\caption{\label{fig:images}Normalized density plots simulating images that appear on the image sensor of a diffraction-limited 1-m optical telescope, at the wavelength $\lambda=1~\mu$m, at 1,000~AU from the Sun, at different positions with respect to the SGL optical axis. Left: The telescope is positioned on the optical axis, $\rho_0=0$, in accordance with Eq.~(\ref{eq:amp-w-f2dp}). Center: The telescope is positioned several meters away from the optical axis in the $\phi_0=-\pi/4$ direction, but still in the region of strong interference, in accordance with Eq.~(\ref{eq:Pvdd+}). Right: The telescope is positioned in the region of weak interference, $\rho_0\gtrsim R_\odot$ from the optical axis, with the resulting minor and major images (their positions swapped, as expected, by the optical telescope) shown in accordance with Eq.~(\ref{eq:FI-ir+}). The Sun is indicated with a dashed (yellow) line, while the Einstein ring is shown as a solid line.
}
\end{figure}

Substituting this result in (\ref{eq:amp-w-d*}), we obtain the amplitude of the EM wave on the focal plane of the optical telescope:
{}
\begin{eqnarray}
{\cal A}({\vec x}_i)&=&i\sqrt{\mu_0}\Big(\frac{kd^2}{8 f}\Big) \bigg(\frac{2}{(\alpha^2-\eta_i^2){\textstyle\frac{1}{2}}d}
\Big( \alpha J_0(\eta_i {\textstyle\frac{1}{2}}d) J_1(\alpha {\textstyle\frac{1}{2}}d)-\eta_i J_0(\alpha {\textstyle\frac{1}{2}}d) J_1(\eta_i {\textstyle\frac{1}{2}}d)\Big)+{\cal O}\big(\alpha\rho_0\big)\bigg)e^{ikf(1+{{\vec p}^2}/{2f^2})}.
  \label{eq:ampD}
\end{eqnarray}
To evaluate the intensity distribution corresponding to the EM signal deposited in the optical telescope's focal plane, we use (\ref{eq:ampD}) in (\ref{eq:Pv}). After averaging over time, we get the time-averaged Poynting vector (i.e., intensity) in the focal plane of the optical telescope, given to ${\cal O}\big(\alpha^2\rho^2_0\big)$ as:
{}
\begin{eqnarray}
S_z(\rho_i)=\frac{cE_0^2}{8\pi}
\Big(\frac{kd^2}{8  f}\Big)^2\mu_0
\bigg(\frac{2}{(\alpha^2-\eta_i^2){\textstyle\frac{1}{2}}d}
 \Big(
 \alpha J_0(\eta_i {\textstyle\frac{1}{2}}d) J_1(\alpha {\textstyle\frac{1}{2}}d)-\eta_i J_0(\alpha {\textstyle\frac{1}{2}}d) J_1(\eta_i {\textstyle\frac{1}{2}}d)\Big)\bigg)^2.
  \label{eq:amp-w-f2dp}
\end{eqnarray}
Expression (\ref{eq:amp-w-f2dp}) is always finite, reaching its maximum  at the Einstein ring for which  $\eta_i=\alpha$,  see discussion in \cite{Turyshev-Toth:2019-extend}. Fig.~\ref{fig:images}~(left) shows the intensity distribution  corresponding to the result (\ref{eq:amp-w-f2dp}).

Result (\ref{eq:amp-w-f2dp}) extends previously known results on the case of imaging with the SGL. In fact, taking the limit of $r_g \rightarrow 0$ (or, equivalently $\alpha \rightarrow0$) in (\ref{eq:amp-w-f2dp}), and remembering the definitions of $\mu_0$ from (\ref{eq:mu}) and of $\alpha$ and $\eta_i$ from (\ref{eq:alpha-mu}), we obtain the Poynting vector that shows the classic Airy pattern characterizing the optical telescope:
{}
\begin{eqnarray}
S^0_z(\rho_i)=\frac{cE_0^2}{8\pi}
\Big(\frac{kd^2}{8f}\Big)^2
\bigg(\frac{2J_1\big( {\textstyle\frac{1}{2}}kd \frac{\rho_i}{f}\big)}{{\textstyle\frac{1}{2}}kd\frac{\rho_i}{f}}\bigg)^2.
  \label{eq:amp-w-f2d2p}
\end{eqnarray}

We note that (\ref{eq:amp-w-f2dp}) is also finite when $\rho_i=0$ (or, equivalently, $\eta_i=0$), with the corresponding value computed as
{}
\begin{eqnarray}
S_z(0)&=&\frac{cE_0^2}{8\pi}
\Big(\frac{kd^2}{8  f}\Big)^2\mu_0
\bigg(\frac{2J_1\Big({\textstyle\frac{1}{2}}kd \sqrt{\frac{2r_g}{\overline z}}\Big)}{{\textstyle\frac{1}{2}}kd \sqrt{\frac{2r_g}{\overline z}}}\bigg)^2.
  \label{eq:amp-w-f2dp7}
\end{eqnarray}
From (\ref{eq:amp-w-f2dp7}) we see that, for $r_g\not=0$ and $\alpha \rho_0\ll1$, the amplification factor at the center of the telescope's focal plane is evaluated to be (with an approximation offered for $d={\cal O}(1~{\rm m})$):
{}
\begin{eqnarray}
\mu^0_{\tt det}=\mu_0\bigg(\frac{2J_1\Big({\textstyle\frac{1}{2}}kd \sqrt{\frac{2r_g}{\overline z}}\Big)}{{\textstyle\frac{1}{2}}kd \sqrt{\frac{2r_g}{\overline z}}}\bigg)^2\simeq
2.02\times 10^7\,\Big(\frac{\lambda}{1\,\mu{\rm m}}\Big)^2\Big(\frac{{\overline z}}{650\,{\rm AU}}\Big)^{\frac{3}{2}}\Big(\frac{1\,{\rm m}}{d}\Big)^3,
  \label{eq:amp-w-f2dps2}
\end{eqnarray}
representing a gravitationally-induced bright spot in the focal plane of the optical telescope, the intensity of which is determined by the wavelength, telescope aperture and distance from the Sun.

For $r_g\rightarrow 0$,  the amplification factor $\mu^0_{\tt det}$ reduces to $\mu^0_{\tt det}=1$ and the result (\ref{eq:amp-w-f2dp7}) is equivalent to (\ref{eq:amp-w-f2d2p}) developed for $\rho_i=0$. One may show that expression (\ref{eq:amp-w-f2dp}) is always finite. In fact, even when $\alpha-\eta_i=k(\sqrt{{2r_g}/{\overline z}}-{\rho_i}/{f})=0$, it remains finite, describing the Einstein ring as seen by the optical telescope, shown in Fig.~\ref{fig:images}~(left), at the position given by
{}
\begin{eqnarray}
\rho_i=f\sqrt{\frac{2r_g}{\overline z}}.
  \label{eq:amp-det}
\end{eqnarray}

Equation~(\ref{eq:amp-w-f2dp}) describes an Einstein ring that is formed in the focal plane of the optical telescope. We demonstrate this by  taking the limit $\eta_i\rightarrow\alpha$ in (\ref{eq:amp-w-f2dp}), we derive the Poynting vector at the Einstein ring:
{}
\begin{eqnarray}
S_z(\rho^{\tt ER}_i)=\frac{cE_0^2}{8\pi}
\Big(\frac{kd^2}{8  f}\Big)^2\mu_0
 \bigg(J^2_0\Big({\textstyle\frac{1}{2}}kd \sqrt{\frac{2r_g}{\overline z}}\Big)+J^2_1\Big({\textstyle\frac{1}{2}}kd \sqrt{\frac{2r_g}{\overline z}}\Big)\bigg)^2.
  \label{eq:amp-w-f2dp+}
\end{eqnarray}

We can derive the intensity distribution at the Einstein ring seen in the optical telescope's focal plane. To do that, we simplify expression (\ref{eq:amp-w-f2dp+}), by using the approximations for the Bessel functions for large arguments \cite{Abramovitz-Stegun:1965}:
\begin{eqnarray}
J_0(x)\simeq \sqrt{\frac{2}{\pi x}}\cos(x-{\textstyle\frac{\pi}{4}})+{\cal O}\big(x^{-1}\big)
\qquad {\rm and} \qquad
J_1(x)\simeq \sqrt{\frac{2}{\pi x}}\sin(x-{\textstyle\frac{\pi}{4}})+{\cal O}\big(x^{-1}\big),
\label{eq:BF}
\end{eqnarray}
which allow us to express (\ref{eq:amp-w-f2dp}), with $\mu_0$ from (\ref{eq:mu}), as
{}
\begin{eqnarray}
S_z(\rho_i^{\tt ER})&=&\frac{cE_0^2}{8\pi}
\Big(\frac{kd^2}{8  f}\Big)^2\Big(\frac{8\lambda \overline z}{\pi^2d^2}\Big).
  \label{eq:amp-w-f2dp*}
\end{eqnarray}

Comparing this expression to (\ref{eq:amp-w-f2d2p}), we see that the light on the Einstein ring is amplified by the factor
{}
\begin{eqnarray}
\mu_{\tt ER}=\frac{8\lambda {\overline z}}{\pi^2 d^2}=7.88\times 10^7\,\Big(\frac{\lambda}{1\,\mu{\rm m}}\Big)\Big(\frac{{\overline z}}{650\,{\rm AU}}\Big)\Big(\frac{1\,{\rm m}}{d}\Big)^2,
  \label{eq:amp-w-f2dps1}
\end{eqnarray}
which is independent of $r_g$, as it is already accounted for by the position of the Einstein ring on the detector (\ref{eq:amp-det}).

\subsection{Large displacements from the optical axis}
\label{sec:large-disp}

Next, we consider the case when the telescope is positioned at a large distance from the optical axis, $\rho_0\gg d$. To compute ${\cal A}({\vec x}_i,{\vec x}_0)$ in (\ref{eq:amp-w-d*}), we note that in this region $\alpha\rho_0\gg 1$, and thus, the Bessel function, $J_0$ can approximated using (\ref{eq:BF}), which results in
 {}
\begin{eqnarray}
  J_0(\alpha |{\vec x}+{\vec x}_0|)
=\frac{1}{\sqrt{2\pi \alpha |{\vec x}+{\vec x}_0|}}\Big(e^{i(\alpha |{\vec x}+{\vec x}_0|-\frac{\pi}{4})}+e^{-i(\alpha |{\vec x}+{\vec x}_0|-\frac{\pi}{4})}\Big).
  \label{eq:bf0}
\end{eqnarray}

Taking into account the fact that in this region $|{\vec x}|\ll \rho_0$,  we expand $ |{\vec x}+{\vec x}_0|$ to first order in ${\vec x}$:
 {}
\begin{eqnarray}
 |{\vec x}+{\vec x}_0|=\rho_0+({\vec x}\cdot {\vec n}_0) +{\cal O}(\rho^2),
  \label{eq:mod}
\end{eqnarray}
where ${\vec n}_0={\vec x}_0/\rho_0$ is the unit vector along ${\vec x}_0$, as given by  (\ref{eq:coord'}). With (\ref{eq:mod}), we may present (\ref{eq:bf0}) as
 {}
\begin{eqnarray}
  J_0\Big(\alpha |{\vec x}+{\vec x}_0|\Big)
=\frac{1}{\sqrt{2\pi \alpha \rho_0}}\Big\{\Big(1-\frac{1}{\rho_0}({\vec x}\cdot{\vec n_0})\Big)\Big(e^{i(\alpha (\rho_0+({\vec x}\cdot{\vec n}_0))-\frac{\pi}{4})}+e^{-i(\alpha (\rho_0+({\vec x}\cdot{\vec n}_0))-\frac{\pi}{4})}\Big) +{\cal O}\Big(\frac{\rho^2}{\rho^2_0}\Big)\Big\}.
  \label{eq:bf}
\end{eqnarray}

As a result, using the definitions (\ref{eq:coord'})--(\ref{eq:coord}) and trigonometrical identities, the double integral in (\ref{eq:amp-w-d*}) takes the form
{}
\begin{eqnarray}
I({\vec x}_i,{\vec x}_0)&=& \iint\displaylimits_{|{\vec x}|^2\leq (d/2)^2}
  d^2{\vec x}
  J_0(\alpha|{\vec x}+{\vec x}_0|) e^{-i\eta_i({\vec x}\cdot{\vec n_i})}=
   \nonumber\\
   &=&
\frac{1}{\sqrt{2\pi \alpha \rho_0}} \Big\{
  \int_0^{2\pi} \hskip -8pt d\phi \int_0^{d/2} \hskip -8pt \rho d\rho \,
  \Big(1-\frac{\rho\cos(\phi-\phi_0)}{\rho_0}\Big)
\Big(e^{i\varphi_+({\vec x})}+e^{i\varphi_-({\vec x})}\Big)+{\cal O}\Big(\frac{\rho^2}{r_0^2}\Big)\Big\},
  \label{eq:am-3*2*}
\end{eqnarray}
where the phases $\varphi_\pm({\vec x})$ are
{}
\begin{eqnarray}
\varphi_\pm({\vec x})&=&
\pm(\alpha \rho_0-{\textstyle\frac{\pi}{4}})+u_\pm\,\rho\, \cos\big(\phi-\epsilon_\pm\big)+{\cal O}(\rho^2),
  \label{eq:ph4}
\end{eqnarray}
with the quantities $u_\pm$ and $\epsilon_\pm$ are given by the following relationships:
{}
\begin{eqnarray}
u_\pm=\sqrt{\alpha^2\mp2\alpha\eta_i\cos\big(\phi_i-\phi_0\big)+\eta_i^2},
  \label{eq:upm}
\qquad
\cos\epsilon_\pm=\frac{\pm\alpha  \cos\phi_0-\eta_i\cos\phi_i}{u_\pm},
\qquad
\sin\epsilon_\pm=\frac{\pm\alpha  \sin\phi_0-\eta_i\sin\phi_i}{u_\pm}.
  \label{eq:eps}
\end{eqnarray}

With this parameterization, the integral (\ref{eq:am-3*2*}) is easy to evaluate:
{}
\begin{eqnarray}
I({\vec x}_i,{\vec x}_0)&=&
\pi \Big(\frac{d}{2}\Big)^2
\frac{1}{\sqrt{2\pi\alpha \rho_0}}  \Big\{e^{i\big(\alpha \rho_0-{\textstyle\frac{\pi}{4}}\big)} \Big(\frac{
2J_1(u_+\frac{1}{2}d)}{u_+ \frac{1}{2}d}-\frac{id}{2\rho_0}\cos(\phi_0-\epsilon_+)\frac{
2J_2(u_+\frac{1}{2}d)}{u_+\frac{1}{2}d}\Big)+\nonumber\\
&&\hskip 60pt+\,
e^{-i\big(\alpha \rho_0-{\textstyle\frac{\pi}{4}}\big)} \Big(\frac{
2J_1(u_-\frac{1}{2}d)}{u_-\frac{1}{2}d}-\frac{id}{2\rho_0}\cos(\phi_0-\epsilon_-)\frac{
2J_2(u_-\frac{1}{2}d)}{u_-\frac{1}{2}d}\Big)+{\cal O}\Big(\frac{d^2}{\rho_0^2}\Big)\Big\}.
  \label{eq:I12**}
\end{eqnarray}

With this result, the complex amplitude of the EM field (\ref{eq:amp-w-d*}) takes the form
{}
\begin{eqnarray}
{\cal A}({\vec x}_i,{\vec x}_0)&=&\sqrt{\mu_0}
{E}_0 \frac{1}{ \sqrt{2\pi\alpha \rho_0}}  \Big(\frac{kd^2}{8f}\Big) e^{i\big(kf(1+{{ p}^2}/{2f^2})+\frac{\pi}{2}\big)}\times\nonumber\\
&&\times
\Big\{e^{i\big(\alpha \rho_0-{\textstyle\frac{\pi}{4}}\big)} \Big(\frac{
2J_1(u_+\frac{1}{2}d)}{u_+ \frac{1}{2}d}-\frac{id}{2\rho_0}\cos(\phi_0-\epsilon_+)\frac{
2J_2(u_+\frac{1}{2}d)}{u_+\frac{1}{2}d}\Big)+\nonumber\\
&&\hskip 60pt+\,
e^{-i\big(\alpha \rho_0-{\textstyle\frac{\pi}{4}}\big)} \Big(\frac{
2J_1(u_-\frac{1}{2}d)}{u_-\frac{1}{2}d}-\frac{id}{2\rho_0}\cos(\phi_0-\epsilon_-)\frac{
2J_2(u_-\frac{1}{2}d)}{u_-\frac{1}{2}d}\Big)+{\cal O}\Big(\frac{d^2}{\rho_0^2}\Big)\Big\}.~~~
  \label{eq:amp-amp}
\end{eqnarray}

Substituting this expression for the complex amplitude in (\ref{eq:amp-w-d*}) and then into (\ref{eq:DB-sol-rho2})--(\ref{eq:Pv}), after time averaging, we obtain the following expression for the Poynting vector on the image plane:
 {}
\begin{eqnarray}
S_z({\vec x}_i,{\vec x}_0)&=&
\frac{cE_0^2}{8\pi}
 \Big(\frac{kd^2}{8f}\Big)^2  \frac{\sqrt{2r_g{\overline z}}}{2\rho_0}  \times \nonumber\\
 &\times&
\Big\{ \Big(\frac{
2J_1(u_+\frac{1}{2}d)}{u_+\frac{1}{2}d}\Big)^2+
\Big(\frac{
2J_1(u_-\frac{1}{2}d)}{u_-\frac{1}{2}d}\Big)^2+2\sin\big(2\alpha \rho_0\big)\frac{
2J_1(u_+\frac{1}{2}d)}{u_+\frac{1}{2}d} \frac{
2J_1(u_-\frac{1}{2}d)}{u_-\frac{1}{2}d}-\nonumber\\
&&\hskip 80pt-\,
 \frac{d}{\rho_0}\cos(2\alpha \rho_0)\Big(\frac{\alpha-\eta_i\cos(\phi_i-\phi_0)}{u_+}\frac{
2J_1(u_-\frac{1}{2}d)}{u_-\frac{1}{2}d}
\frac{
2J_2(u_+\frac{1}{2}d)}{u_+ \frac{1}{2}d}+\nonumber\\
&&\hskip 150pt+\,
\frac{\alpha+\eta_i\cos(\phi_--\phi_0)}{u_-}\frac{
2J_1(u_+\frac{1}{2}d)}{u_+\frac{1}{2}d}
\frac{
2J_2(u_-\frac{1}{2}d)}{u_- \frac{1}{2}d}\Big)+{\cal O}\Big(\frac{d^2}{\rho_0^2}\Big)
\Big\},~~~
  \label{eq:Pvdd}
\end{eqnarray}
where we used the definitions for $\mu_0$, $\alpha$ and $\eta_i$ from (\ref{eq:mu}) and (\ref{eq:alpha-mu}) as well as for $u_\pm$ and $\epsilon_\pm$  from (\ref{eq:upm}).

We observe that the ratios involving the Bessel functions in the expression (\ref{eq:Pvdd}) are at most $2J_1(x)/x=1$, and only for $x=0$. Given the fact that the spatial frequency $\alpha$ is quite high, for any other value of the argument these ratios are negligibly small. In addition, the last term in this expression is at most $\propto d/\rho_0$, which is negligibly small even compared to the smallest term (i.e., mixed containing $\sin2\alpha\rho_0$). Therefore, the last terms in this expression may be neglected, allowing us to present a simplified form of Eq.~(\ref{eq:Pvdd}):
{}
\begin{eqnarray}
S_z({\vec x}_i,{\vec x}_0)&=&
\frac{cE_0^2}{8\pi}
 \Big(\frac{kd^2}{8f}\Big)^2  \frac{\sqrt{2r_g{\overline z}}}{2\rho_0}
\Big\{ \Big(\frac{2J_1(u_+\frac{1}{2}d)}{u_+\frac{1}{2}d}\Big)^2+
\Big(\frac{2J_1(u_-\frac{1}{2}d)}{u_-\frac{1}{2}d}\Big)^2+{\cal O}\Big(\frac{d^2}{\rho_0^2}\Big)
\Big\}.
  \label{eq:Pvdd+}
\end{eqnarray}

Equation (\ref{eq:Pvdd+}) describes two spots of light of nearly equal intensity, as shown in Fig.~\ref{fig:images}~(center). This is what remains from the Einstein ring as the telescope is displaced at a large distance from the optical axis, but still staying within the strong interference region of the SGL.

Remembering the definition for $u_\pm$ from (\ref{eq:eps}) and taking the limit $\eta_i\rightarrow \alpha$, we obtain an expression for the Poynting vector at the Einstein ring to the order of ${\cal O}({d^2}/{\rho_0^2})$:
{}
\begin{eqnarray}
S_z(\rho_i^{\tt ER},\phi_i,{\vec x}_0)&=&
\frac{cE_0^2}{8\pi}
 \Big(\frac{kd^2}{8f}\Big)^2  \frac{\sqrt{2r_g{\overline z}}}{2\rho_0}
\Big\{ \Big(\frac{2J_1\big(\alpha d\sin{\textstyle\frac{1}{2}}(\phi_i-\phi_0)\big)}{\alpha d\sin{\textstyle\frac{1}{2}}(\phi_i-\phi_0)}\Big)^2+
\Big(\frac{2J_1\big(\alpha d\cos{\textstyle\frac{1}{2}}(\phi_i-\phi_0)\big)}{\alpha d\cos{\textstyle\frac{1}{2}}(\phi_i-\phi_0)}\Big)^2
\Big\}.
  \label{eq:Pvdd+*}
\end{eqnarray}
Given the fact that the product $\alpha d$ is quite large, Eq.~(\ref{eq:Pvdd+*}) is close to zero everywhere except for two peaks where the arguments of the two Bessel functions vanish. This vanishing depends on the direction of the displacement from the optical axis, $\phi_0$. As a result,  (\ref{eq:Pvdd+*}) describes  two peaks that  appear in the optical telescope's focal plane, at the same radial distance $\rho_i=\rho_i^{\tt ER}$,  but in opposite directions, which are given as $\phi_i=\phi_0$ and $\phi_i=\phi_0+\pi$.

\section{Imaging in the geometric optics and weak interference regions}
\label{sec:go-wint}

If we position the telescope further away from the optical axis, it enters the weak interference region of the SGL (see Fig.~\ref{fig:regions}), where for any given point source at any point on the image plane, two rays of light are present, corresponding to the incident and scattered wave \cite{Turyshev-Toth:2017}. Moving still further from the optical axis, the telescope enters the geometric optics region of the SGL where at any given point only one ray of light is present, with the other being blocked by the Sun.

Although a description of the optical properties of the SGL in these regions is of little practical importance insofar as imaging of distant sources is concerned, it can provide a wave-optical description of microlensing phenomena. Such a wave-optical description is still largely absent in ongoing microlensing modeling efforts \cite{Liebes:1964,Refsdal:1964,Schneider-Ehlers-Falco:1992}. Below, we describe the relevant EM fields in these two regions and derive the intensity distribution pattern in the focal plane of an imaging telescope.

\subsection{The EM field in the geometric optics and weak interference regions}

In Ref.~\cite{Turyshev-Toth:2019-extend}, we considered a high-frequency EM wave (i.e., neglecting terms $\propto(kr)^{-1}$) and for $r\gg r_g$ and derived the components of the EM field in the geometric optics and weak interference regions. As light amplification in these regions is rather weak, it is sufficient to derive the solution to the highest leading order term in the image field amplitude. However, as the wavenumber for optical wavelengths, $k$, is rather large, the phase of the resulting solution must include all the relevant terms.

For a source at a distance $r_0$ from the Sun, the components of the EM field needed to estimated the flux through the image plane can be given to the required order in the spherical coordinate system in the following form:
{}
\begin{eqnarray}
    \left( \begin{aligned}
{D}_\theta& \\
{B}_\theta& \\
  \end{aligned} \right) =    \left( \begin{aligned}
{B}_\phi& \\
-{D}_\phi& \\
  \end{aligned} \right)&=&
 \left( \begin{aligned}
 \cos\phi& \\
 \sin\phi& \\
  \end{aligned} \right)\,e^{-i\omega t}\gamma(r, \theta)+{\cal O}(r_g^2, \theta^2, b/z_0).
  \label{eq:DB-sol-rho_go}
\end{eqnarray}
The term $\gamma(r, \theta)$, for large partial momenta, $\ell\gg1$,  following \cite{Turyshev-Toth:2019-extend}, is determined from the following integral:

{}
\begin{eqnarray}
\gamma(r, \theta)&=&
E_0u
 \frac{e^{ik(r+r_0+r_g\ln 4k^2rr_0)}}{kr} \int_{\ell=kR_\odot^\star}^\infty \hskip -5pt
\frac{\sqrt{\ell}d\ell}{\sqrt{2\pi\sin\theta}}e^{i\big(2\sigma_\ell+{\ell^2}/{2k\tilde r}\big)}
\Big(e^{i(\ell\theta+{\textstyle\frac{\pi}{4}})}-e^{-i(\ell\theta+{\textstyle\frac{\pi}{4}})}\Big),
  \label{eq:gamma**1*}
\end{eqnarray}
where $1/\tilde r=1/r+1/r_0$. The radial components of the EM wave behave as $({D}_r, {B}_r)\sim {\cal O}({\rho}/{z},b/z_0)$ and, thus, they are negligibly small compared to the other two components (\ref{eq:DB-sol-rho_go}).

As was done in \cite{Turyshev-Toth:2019-extend}, we evaluate this integral by the method of stationary phase. To do that, we see that the relevant $\ell$-dependent part of the phase in (\ref{eq:gamma**1*})  is of the form
{}
\begin{equation}
\varphi_{\pm}(\ell)=\pm\big(\ell\theta+\textstyle{\frac{\pi}{4}}\big)+2\sigma_\ell +\dfrac{\ell^2}{2k \tilde r}+{\cal O}\big(r_g^2, (kr)^{-3}\big),
\label{eq:S-l}
\end{equation}
where for $\ell\gg kr_g$  the Coulomb phase shift, $\sigma_\ell$, has the from: $\sigma_\ell= -kr_g\ln \ell.$ The phase is stationary when $d\varphi_{\pm}/d\ell=0$, which implies
{}
\begin{equation}
\pm\theta-\frac{2r_g}{b}+\frac{b}{\tilde r}={\cal O}\big(r_g^2, (kr)^{-3}\big),
\label{eq:S-l-pri=}
\end{equation}
where we used the semiclassical relationship between the partial momentum $\ell$ and the impact parameter $b$, given as $\ell\simeq kb$. This quadratic equation yields two families of solutions:
{}
\begin{equation}
b_{\tt in}= \mp {\textstyle\frac{1}{2}} \Big(\tilde r \theta+\sqrt{\tilde r^2 \theta^2+8r_g \tilde r}\Big)+{\cal O}(\theta^3,r_g^2), \qquad {\rm and} \qquad
b_{\tt sc}= \mp {\textstyle\frac{1}{2}} \Big(\tilde r \theta-\sqrt{\tilde r^2 \theta^2+8r_g \tilde r}\Big)+{\cal O}(\theta^3,r_g^2),
\label{eq:S-l-pri}
\end{equation}
where $b_{\tt in}$ and $b_{\tt sc}$ are two families of impact parameters describing incident and scattered EM waves, corresponding to light rays passing by the near side and the far side of the Sun (with respect to the location of the telescope), correspondingly. After it is diffracted by a point-source gravitational lens, a wavefront is described as the sum of a gravity-modified plane wave (the incident wave) and a spherical wave centered on the gravitational lensing source (the scattered wave);
see, for instance, Fig.~2 of \cite{Turyshev-Toth:2017}.  The impact parameters (\ref{eq:S-l-pri}) correspond to images that appear close to the Einstein ring on opposite sides of the lens; the ``scattered'' image, denoted by ``${\tt sc}$'', on the far side relative to the telescope (called the minor image) always appears inside the Einstein ring, and the ``incident'' image, denoted by ``${\tt in}$'' on the near side always appears outside (major image, see \cite{Schneider-Ehlers-Falco:1992} for details).

For $\theta\gg \sqrt{2r_g/\tilde r}$, our result is equivalent to the two solutions derived  in Sec.~IV of  \cite{Turyshev-Toth:2019-extend}. However, the form (\ref{eq:S-l-pri}) allows us to study the behavior  of the EM wave in the transition between the two solutions in the region where angle $\theta$ is of the same order as the Einstein deflection angle $\theta\sim \sqrt{2r_g/\tilde r}$.

By dividing the solutions (\ref{eq:S-l-pri}) by $\tilde r$, we may present them in term of the angles $\theta_+=b_{\tt in}/\tilde r$ and $\theta_-=b_{\tt in}/\tilde r$:
{}
\begin{equation}
\theta_+= {\textstyle\frac{1}{2}} \Big(\sqrt{\theta^2+4\theta_E^2}+\theta\Big), \qquad {\rm and} \qquad
\theta_-= -{\textstyle\frac{1}{2}} \Big(\sqrt{\theta^2+4\theta_E^2}-\theta\Big),
\label{eq:S-l-pri-mic}
\end{equation}
where $\theta_E=\sqrt{{2r_g}/{\tilde r}}$ is the Einstein deflection angle.  This establishes the correspondence of our analysis in this section to the well-known modeling of microlensing \cite{Liebes:1964,Refsdal:1964,Schneider-Ehlers-Falco:1992}. Expressions (\ref{eq:S-l-pri-mic})  lead to the familiar expression to describe the image magnification of $A=(u^2+2)/(u\sqrt{u^2+4})$, where $u=\theta/\theta_E$.  Our description allows us to develop the vectorial description of the microlensing phenomena and, besides magnification, it also allows us to describe light amplification.

Following the approach presented in  \cite{Turyshev-Toth:2019-extend}, we again use the method of stationary phase (\ref{eq:S-l}) for the first family of solutions of (\ref{eq:S-l-pri}), corresponding to $\ell_{\tt in}=kb_{\tt in}$. This results in the factor $\gamma_{\tt in}(r,\theta)$ corresponding to the incident EM wave moving towards the interference region:
{}
\begin{eqnarray}
\gamma_{\tt in}(r,\theta)&=&
E_0\, a_{\tt in}(\tilde r,\theta)
e^{i\big(k(r+r_0+r_g\ln 4k^2rr_0)+\varphi_{\tt in}
(\tilde r,\theta)\big)}+{\cal O}(\theta^4, \frac{r_g}{r}\theta^2),
\label{eq:Pi_s_exp4+1pp}\\
a^2_{\tt in}(\tilde r,\theta)&=&\frac{\big({\textstyle\frac{1}{2}} (\sqrt{1+{8r_g}/{\tilde r\theta^2}}+1)\big)^2}{\sqrt{1+{8r_g}/{\tilde r\theta^2}}},
\nonumber\\
\varphi_{\tt in}(\tilde r,\theta)
&=&-k\Big({\textstyle\frac{1}{4}} \theta \big(\tilde r \theta+\sqrt{\tilde r^2 \theta^2+8r_g \tilde r}\big)-r_g+2r_g\ln{\textstyle\frac{1}{2}} k \big(\tilde r \theta+\sqrt{\tilde r^2 \theta^2+8r_g \tilde r}\big)\Big).
\nonumber
\end{eqnarray}

Next, we consider the second family of solutions in (\ref{eq:S-l-pri}), given by $\ell_{\tt sc}=kb_{\tt sc}$. It allows us to compute the $\gamma_{\tt sc}(r,\theta)$
factor for the scattered wave, given as
{}
\begin{eqnarray}
\gamma_{\tt sc}(r,\theta)&=&
E_0
\, a_{\tt sc}(\tilde r,\theta)e^{i\big(k(r+r_0+r_g\ln 4k^2rr_0)+\varphi_{\tt sc}
(\tilde r,\theta)\big)}
+{\cal O}(\theta^2, \frac{r_g}{r}\theta^2),
\label{eq:gamma-2}
\\
a^2_{\tt sc}(\tilde r,\theta)&=&\frac{\big({\textstyle\frac{1}{2}} (\sqrt{1+{8r_g}/{\tilde r\theta^2}}-1)\big)^2}{\sqrt{1+{8r_g}/{\tilde r\theta^2}}},
\nonumber\\
\varphi_{\tt sc}
(\tilde r,\theta)&=&-k\Big({\textstyle\frac{1}{4}} \theta \big(\tilde r \theta-\sqrt{\tilde r^2 \theta^2+8r_g \tilde r}\big)-r_g+2r_g\ln {\textstyle\frac{1}{2}} k \big(\tilde r \theta-\sqrt{\tilde r^2 \theta^2+8r_g \tilde r}\big)\Big). \nonumber
\end{eqnarray}

As a result, the components of the incident EM field to the order of ${\cal O}\big(r_g^2, \theta^2, b/z_0\big)$ take the form
{}
\begin{eqnarray}
    \left( \begin{aligned}
{D}_\theta& \\
{B}_\theta& \\
  \end{aligned} \right)_{\tt \hskip -2pt in/sc}
=    \left( \begin{aligned}
{B}_\phi& \\
-{D}_\phi& \\
  \end{aligned} \right)_{\tt \hskip -2pt in/sc}&=&
 E_0
  {\cal A}_{\tt in/sc}
  (\tilde r,\theta) e^{i\big(k(r+r_0+r_g\ln 4k^2rr_0)-\omega t\big)}
 \left( \begin{aligned}
 \cos\phi& \\
 \sin\phi& \\
  \end{aligned} \right),
  \label{eq:DB-sol-in}
\end{eqnarray}
with the complex amplitudes ${\cal A}_{\tt in}$ and ${\cal A}_{\tt sc}$ given as
{}
\begin{eqnarray}
{\cal A}_{\tt in}(\tilde r,\theta)&=&
a_{\tt in}(\tilde r,\theta)
\exp\Big[{-ik\Big\{{\textstyle\frac{1}{4}} \theta \big(\tilde r \theta+\sqrt{\tilde r^2 \theta^2+8r_g \tilde r}\big)\big)-r_g+2r_g\ln {\textstyle\frac{1}{2}} k \big(\tilde r \theta+\sqrt{\tilde r^2 \theta^2+8r_g \tilde r}\big)\Big\}}\Big],
  \label{eq:DB-sol-inA}\\
{\cal A}_{\tt sc}(\tilde r,\theta)&=&
a_{\tt sc}(\tilde r,\theta)
\exp\Big[{-ik\Big\{{\textstyle\frac{1}{4}} \theta \big(\tilde r \theta-\sqrt{\tilde r^2 \theta^2+8r_g \tilde r}\big)\big)-r_g+2r_g\ln {\textstyle\frac{1}{2}} k \big(\tilde r \theta-\sqrt{\tilde r^2 \theta^2+8r_g \tilde r}\big)\Big\}}\Big],
  \label{eq:DB-sol-scA}
\end{eqnarray}
where the $r$-components of the EM waves behave as $({E}_r, {H}_r)_{\tt \hskip 0pt in/sc} \sim {\cal O}({\rho}/{r},b/r_0)$. Note that if $ \theta\gg \sqrt{2r_g/\tilde r}$, results are identical to those reported in \cite{Turyshev-Toth:2019-extend}.

As our concern is the EM field in the image plane, it is convenient to transform these solutions to cylindrical coordinates $(\rho,\phi,z)$, as was done in  \cite{Turyshev-Toth:2017,Turyshev-Toth:2019-extend}. Transforming (\ref{eq:DB-sol-in})  and (\ref{eq:DB-sol-scA}) yields the components of both solutions, to ${\cal O}(r_g^2, \theta^2, b/r_0)$, in the form
{}
\begin{eqnarray}
    \left( \begin{aligned}
{E}_\rho& \\
{H}_\rho& \\
  \end{aligned} \right)_{\tt \hskip -2pt in/sc} =    \left( \begin{aligned}
{H}_\phi& \\
-{E}_\phi& \\
  \end{aligned} \right)_{\tt \hskip -2pt in/sc}&=&
E_0
  {\cal A}_{\tt in/sc}
  \big(\tilde r,\theta\big)e^{i\big(k(r+r_0+r_g\ln k^2rr_0)-\omega t\big)}
 \left( \begin{aligned}
 \cos\overline \phi& \\
 \sin\overline \phi& \\
  \end{aligned} \right),
  \label{eq:DB-sol-in-cc}
\end{eqnarray}
where the $z$-components of the EM waves behave as $({E}_z, {H}_z)_{\tt \hskip 0pt in/sc} \sim {\cal O}({\rho}/{z},\sqrt{2r_gz}/z_0)$, and where $\overline\phi$ is the angle that corresponds to the rotated $\overline z$ coordinate axis described in \cite{Turyshev-Toth:2019-extend}.

Expressing $\tilde r\theta$ via the angle $\beta=b/r_0$, where $b=\sqrt{2r_gz}$ is the impact parameter,  and generalizing the resulting expression to a 3-dimensional case, as was preseted in  \cite{Turyshev-Toth:2019-extend}, for a point source on the optical axis \cite{Turyshev-Toth:2019-extend}, we have
{}
\begin{eqnarray}
\tilde r\theta&=&r\big(\theta+\beta\big)+{\cal O}(r^3/r_0^2)\simeq|\vec{x}+{\vec x}_0|+{\cal O}(r^3/r_0^2) \qquad {\rm and} \qquad \theta\simeq\frac{1}{\overline z}|\vec{x}+{\vec x}_0|.
\label{eq:tildebeta}
\end{eqnarray}
These results allows us to express the complex amplitudes $ {\cal A}_{\tt in/sc}(r,\theta)\rightarrow {\cal A}_{\tt in}({\vec x},{\vec x}_0)$, which is needed for our purposes.

\subsection{Very large displacements from the optical axis}
\label{sec:v-large-disp}

Moving still further away from the optical axis, for angles $\theta\gg\sqrt{2r_g/{\tilde z}}$, we traverse the region of weak interference toward the region of geometric optics.  As discussed in \cite{Turyshev-Toth:2019-extend}, in the region of the geometric optics  at any given point on the image plane we have only
the incident ray of light, which is the ray passing on the near side of the Sun with respect to the telescope; the scattered ray on the opposite side is blocked by the opaque sphere of the Sun \cite{Turyshev-Toth:2019-extend}.
However, in the region of weak interference at any given point on the image plane
both rays are still present
\cite{Turyshev-Toth:2017,Turyshev-Toth:2019-extend}.
Based on (\ref{eq:DB-sol-in-cc}), the incident and scattered EM waves
on the image plane are given, in cylindrical coordinates $(\rho,\phi,z)$, as
{}
\begin{eqnarray}
    \left( \begin{aligned}
{E}_\rho& \\
{H}_\rho& \\
  \end{aligned} \right)_{\tt \hskip -2pt in/sc} =    \left( \begin{aligned}
{H}_\phi& \\
-{E}_\phi& \\
  \end{aligned} \right)_{\tt \hskip -2pt in/sc}&=&
E_0
  {\cal A}_{\tt in/sc} \big({\vec x},{\vec x}_0\big)
  e^{i\big(k(r+r_0+r_g\ln k^2rr_0)-\omega t\big)}
 \left( \begin{aligned}
\cos\overline \phi& \\
\sin\overline \phi& \\
  \end{aligned} \right),
  \label{eq:DB-pol-sol}
\end{eqnarray}
with the complex amplitudes  ${\cal A}_{\tt in/sc} ({\vec x},{\vec x}_0)$ for the incident and scattered waves from (\ref{eq:DB-sol-inA})--(\ref{eq:DB-sol-scA}), correspondingly. The $z$-components of the EM waves behave as $({E}_z, {H}_z)_{\tt \hskip 0pt in/sc} \sim {\cal O}({\rho}/{z},b/z_0)$.

As in this case $\rho \ll \rho_0$, we may use the approximation given in (\ref{eq:mod}), which allows us to expand (\ref{eq:DB-sol-inA}) and (\ref{eq:DB-sol-scA}),
to terms first order in $\rho/r_0$, yielding the following results:
{}
\begin{eqnarray}
{\cal A}_{\tt in}({\vec x},{\vec x}_0)&=&
a_{\tt in}(\rho_0,\tilde r)
\exp\Big(i\delta\varphi_{\tt in}(\rho_0,\tilde r)-i\big(\xi_{\tt in}({\vec x}\cdot{\vec n}_0)+\eta_i({\vec x}\cdot{\vec n}_i)\big)\Big),
  \label{eq:amp-Ain}\\
  {\cal A}_{\tt sc}
  ({\vec x},{\vec x}_0)&=&
  a_{\tt sc}(\rho_0,\tilde r)\exp\Big(i\delta\varphi_{\tt sc}
  (\rho_0,\tilde r)+{i\big(\xi_{\tt sc}({\vec x}\cdot{\vec n}_0)-\eta_i({\vec x}\cdot{\vec n}_i)\big)}\Big),
  \label{eq:amp-Asc}
\end{eqnarray}
with amplitude factors $a_{\tt in/sc}(\rho_0,{\tilde r})$ and phases $\delta\varphi_{\tt in/sc}(\rho_0,{\tilde r}) $ (with the upper and lower signs for the ``{\tt in}'' and ``{\tt sc}'' waves, correspondingly),  are given as
{}
  \begin{eqnarray}
a^2_{\tt in/sc}(\rho_0,\tilde r) &=&
\frac{\big[{\textstyle\frac{1}{2}} (\sqrt{1+{8r_g\tilde r}/{\rho_0^2}}\pm1)\big]^2}{\sqrt{1+{8r_g\tilde r}/{\rho_0^2}}},
    \label{eq:q_insc}\\
    \delta\varphi_{\tt in/sc}
(\rho_0,\tilde r) &=& -k\Big(\frac{\rho^2_0}{4\tilde r}\Big(1\pm\sqrt{1+\frac{8r_g \tilde r}{\rho_0^2}}-\frac{4r_g \tilde r}{\rho_0^2}\Big)+2r_g\ln k\rho_0{\frac{1}{2}}\Big(\sqrt{1+\frac{8r_g \tilde r}{\rho^2_0}}\pm1\Big)\Big).
    \label{eq:Ain-d_ph}
\end{eqnarray}
We note that when the angles $\theta$ are large,   $\theta \gg \sqrt{2r_g/\tilde r}$,  and thus, $\rho_0\gg \sqrt{2r_g\tilde r}$,  the factors $a_{\tt in/sc}$ in (\ref{eq:q_insc})  take their known values (see \cite{Turyshev-Toth:2019-extend} for details), namely $a^2_{\tt in}(\rho_0,\tilde r)=1+{\cal O}(r_g\theta^2,r_g^2)$ and $a^2_{\tt sc}(\rho_0,\tilde r) =({2r_g\tilde r}/{\rho_0^2})^2 +{\cal O}(r_g\theta^2,r_g^2)$. However, our new expressions (\ref{eq:q_insc}) allow studying the cases when  $\rho_0\simeq \sqrt{2r_g\tilde r}$.

In addition, the spatial frequencies $\xi_{\tt in/sc}$ present in (\ref{eq:amp-Ain})--(\ref{eq:amp-Asc}), are defined by (with $\theta_\pm$ from (\ref{eq:S-l-pri-mic})):
  \begin{eqnarray}
\xi_{\tt in/sc}&=&k\Big(\sqrt{1+\frac{8r_g \tilde r}{\rho^2_0}}\pm1\Big)\frac{\rho_0}{2\tilde r}
\equiv k\theta_\pm.
  \label{eq:betapm}
  \end{eqnarray}
Therefore, to derive  the amplitudes of the EM field in the focal plane of the optical telescope, corresponding to  (\ref{eq:amp-Ain}) and  (\ref{eq:amp-Asc}), we need to put these expressions in  (\ref{eq:amp-w-f}) and evaluate an integral of the type
{}
\begin{eqnarray}
\iint\displaylimits_{|{\vec x}|^2\leq (d/2)^2}\hskip -8pt
  d^2{\vec x}\,e^{-i\big(\xi_{\tt in/sc}({\vec x}\cdot{\vec n}_0)\pm\eta_i({\vec x}\cdot{\vec n}_i)\big)}.
  \label{eq:amp-int}
\end{eqnarray}
This can be done analogously to the derivations in Sec.~\ref{sec:large-disp}. For this we present the phase  in (\ref{eq:amp-int}) as
{}
\begin{eqnarray}
\xi_{\tt in/sc}({\vec x}\cdot{\vec n}_0)\pm\eta_i({\vec x}\cdot{\vec n}_i)=v_\pm\,\rho\, \cos\big(\phi-\sigma_\pm\big)+{\cal O}(\rho^2),
  \label{eq:ph4s}
\end{eqnarray}
where, for convenience, we defined
{}
\begin{eqnarray}
v_\pm&=&\sqrt{\xi_{\tt in/sc}^2\pm2\xi_{\tt in/sc}\eta_i\cos\big(\phi_i-\phi_0\big)+\eta_i^2}, \label{eq:vpms}\\
\cos\sigma_\pm&=&
\frac{\xi_{\tt in/sc}  \cos\phi_0\pm\eta_i\cos\phi_i}{v_\pm},
\qquad
\sin\sigma_\pm=\frac{\xi_{\tt in/sc}  \sin\phi_0\pm\eta_i\sin\phi_i}{v_\pm}.~~
\nonumber
\end{eqnarray}

With these definitions, and using the parameterization given in (\ref{eq:x-im}), the integral (\ref{eq:amp-int}) may be evaluated as
{}
\begin{eqnarray}
  \int_0^{2\pi} \hskip -8pt d\phi \int_0^{d/2} \hskip -8pt \rho d\rho\,
  e^{-iv_\pm\rho\cos(\phi-\sigma_\pm)}=\pi\Big(\frac{d}{2}\Big)^2\, \frac{
2J_1(v_\pm\frac{1}{2}d)}{v_\pm \frac{1}{2}d}.
  \label{eq:amp-int*}
\end{eqnarray}
As a result,  using (\ref{eq:amp-Ain}) and (\ref{eq:amp-Asc}) in  (\ref{eq:amp-w-f})  leads to the following amplitudes of the two EM waves on the optical telescope's image plane:
{}
\begin{eqnarray}
{\cal A}_{\tt in}({\vec x}_i,{\vec x}_0)&=&
\Big(\frac{kd^2}{8f}\Big)\, \Big\{
a_{\tt in}
\Big(\frac{
2J_1(v_+\frac{1}{2}d)}{v_+ \frac{1}{2}d}\Big)e^{i\big(kf(1+{{\vec x}_i^2}/{2f^2})+\delta\varphi_{\tt in}
(\rho_0,\tilde r) +\frac{\pi}{2}\big)}+{\cal O}(r_g^2)\Big\},
  \label{eq:amp-Aind}\\
  {\cal A}_{\tt sc}({\vec x}_i,{\vec x}_0)&=&
\Big(\frac{kd^2}{8f}\Big)\Big\{
a_{\tt sc}
\Big(\frac{
2J_1(v_-\frac{1}{2}d)}{v_- \frac{1}{2}d}\Big)e^{i\big(kf(1+{{\vec x}_i^2}/{2f^2})+\delta\varphi_{\tt sc}
(\rho_0,\tilde r) +\frac{\pi}{2}\big)}+{\cal O}\Big(\frac{r_g\rho_0^2}{{\tilde r}^3}\Big)\Big\}.
  \label{eq:amp-Ascd}
\end{eqnarray}

Remembering the time-dependent phase from (\ref{eq:DB-pol-sol}), we  substitute this expression in (\ref{eq:Pv}) and, after time averaging, we derive the Poynting vector of the EM wave in the focal plane of the imaging telescope.  As a result, in the region of the geometric optics, where only the incident EM wave is present, the intensity of the EM field in the optical telescope's focal plane is derived using (\ref{eq:amp-Aind}), resulting in
 {}
\begin{eqnarray}
S_{\tt geom.o.}({\vec x}_i,{\vec x}_0)&=&
\frac{cE_0^2}{8\pi}
 \Big(\frac{kd^2}{8f}\Big)^2\Big\{a^2_{\tt in}
 \Big(\frac{
2J_1(v_+\frac{1}{2}d)}{v_+\frac{1}{2}d}\Big)^2+{\cal O}(r^2_g)\Big\}.
  \label{eq:FI-go}
\end{eqnarray}
Examining (\ref{eq:vpms}), we see that because the combination $\xi_{\tt in/sc} {\textstyle\frac{1}{2}}d$ may be rather large,  expression (\ref{eq:FI-go}) is almost zero everywhere except for one point where the argument of the Bessel function vanishes. Taking in (\ref{eq:FI-go}) the limit  $\eta_i\rightarrow \xi_{\tt in}$ and considering the case of $2r_g{\tilde r}/\rho_0^2\rightarrow 0$ and taking only the leading term in $a_{\tt in}$, thus taking $a_{\tt in}\rightarrow 1$, we have
 {}
\begin{eqnarray}
S_{\tt geom.o.}({\vec \xi}^{\tt in}_i \hskip -3pt ,{\vec x}_0)&=&
\frac{cE_0^2}{8\pi}
 \Big(\frac{kd^2}{8f}\Big)^2\Big\{
\Big(\frac{
2J_1\big(\xi_{\tt in}d\cos{\textstyle\frac{1}{2}}(\phi_i-\phi_0)\big)}{\xi_{\tt in}d\cos{\textstyle\frac{1}{2}}(\phi_i-\phi_0)}\Big)^2+{\cal O}(r^2_g)\Big\}.
  \label{eq:FI-go+}
\end{eqnarray}
This expression describes one peak corresponding to the incident wave whose intensity is not amplified by the SGL. It is for the image that was derived using  $b_{\tt in}$, corresponding $\xi_{\tt in}$, which always appears outside the Einstein ring.

As in the region of weak interference, both incident and scattered waves are present, the field intensity in the focal plane of the imaging telescope is derived using the sum of the two solutions, (\ref{eq:amp-Aind}) and (\ref{eq:amp-Ascd}), yielding
{}
\begin{eqnarray}
S_{\tt weak.int.}({\vec x}_i,{\vec x}_0)&=&
\frac{cE_0^2}{8\pi}
 \Big(\frac{kd^2}{8f}\Big)^2\Big\{
 a^2_{\tt in}
 \Big(\frac{
2J_1(v_+\frac{1}{2}d)}{v_+\frac{1}{2}d}\Big)^2+
a^2_{\tt sc}
\Big(\frac{
2J_1(v_-\frac{1}{2}d)}{v_-\frac{1}{2}d}\Big)^2+\nonumber\\
&&\hskip -70pt +\,
2\cos\Big(\frac{k\rho_0}{2\tilde r}\sqrt{\rho_0^2+8r_g \tilde r}+2kr_g\ln \frac{\sqrt{\rho^2_0+8r_g \tilde r}+\rho_0}{\sqrt{\rho_0^2+8r_g \tilde r}-\rho_0}\Big)
a_{\tt in}
a_{\tt sc}
\Big(\frac{
2J_1(v_+\frac{1}{2}d)}{v_+\frac{1}{2}d}\Big)\bigg(\frac{
2J_1(v_-\frac{1}{2}d)}{v_-\frac{1}{2}d}\Big)+{\cal O}\Big(\frac{r_g\rho_0^2}{{\tilde r}^3}\Big)\Big\}.
  \label{eq:FI-ir}
\end{eqnarray}

Similar simplifying assumptions based on the behavior of the ratios involving the Bessel function $2J_1(v_\pm\frac{1}{2}d)/{v_\pm\frac{1}{2}d}$ in these regions that led to (\ref{eq:Pvdd+}), are applicable here. Therefore, the intensity distribution pattern in the weak interference region takes the following simplified form
{}
\begin{eqnarray}
S_{\tt weak.int.}({\vec x}_i,{\vec x}_0)&=&
\frac{cE_0^2}{8\pi}
 \Big(\frac{kd^2}{8f}\Big)^2\Big\{
  a^2_{\tt in}
  \Big(\frac{
2J_1(v_+\frac{1}{2}d)}{v_+\frac{1}{2}d}\Big)^2+
 a^2_{\tt sc}
 \Big(\frac{
2J_1(v_-\frac{1}{2}d)}{v_-\frac{1}{2}d}\Big)^2+{\cal O}\Big(\frac{r_g\rho_0^2}{{\tilde r}^3}\Big)\Big\}.
  \label{eq:FI-ir+}
\end{eqnarray}
This expression describes two peaks of uneven brightness, conventionally called major and minor images (see right figure in Fig.~\ref{fig:images}), with the major image depending on $v_+$ from (\ref{eq:vpms})  (and, thus, its behavior is driven by $\theta_+$ characteristic of the incident wave) that appears outside the Einstein ring and the minor image given by the $v_-$-dependent term (and thus on $\theta_-$ corresponding to the scattered wave) that appears inside the Einstein ring. This is the typical behavior observed in the microlensing experiments. The image described by Eq.~(\ref{eq:FI-ir+}) is the inverted image that appears in the focal plane of the convex lens. This behavior is evident in Fig.~\ref{fig:images} (right).

To derive the intensity distribution in the vicinity for the Einstein ring, similarly to (\ref{eq:FI-go+}), we take the limit in $\eta_i\rightarrow \xi_{\tt in/sc}$ in the expression (\ref{eq:FI-ir+}) and again considering the case of $2r_g{\tilde r}/\rho_0^2\rightarrow 0$ and taking only the leading term in $a_{\tt in/sc}$ from (\ref{eq:q_insc}), namely $a_{\tt in}\rightarrow 1$ and $a_{\tt sc}\rightarrow {2r_g{\tilde r}}/{\rho^2_0}$, we  obtain
{}
\begin{eqnarray}
S_{\tt weak.int.}({\vec \xi}^{\tt ER_\pm}_i  \hskip -3pt ,{\vec x}_0)&=&
\frac{cE_0^2}{8\pi}
 \Big(\frac{kd^2}{8f}\Big)^2\Big\{\Big(\frac{
2J_1\big(\xi_{\tt in}d\cos{\textstyle\frac{1}{2}}(\phi_i-\phi_0)\big)}{\xi_{\tt in}d\cos{\textstyle\frac{1}{2}}(\phi_i-\phi_0)\big)}\Big)^2+ \Big(\frac{2r_g{\tilde r}}{\rho^2_0}  \Big)^2 \Big(\frac{
2J_1\big(\xi_{\tt sc}d\sin{\textstyle\frac{1}{2}}(\phi_i-\phi_0)\big)}{\xi_{\tt sc}d\sin{\textstyle\frac{1}{2}}(\phi_i-\phi_0)\big)}\Big)^2
\Big\},~~~
  \label{eq:FI-ir+*}
\end{eqnarray}
with the superscript ${\tt ER}_\pm$ indicates that the two peaks that are located outside and inside of the Einstein ring, correspondingly. Similarly to Eq.~(\ref{eq:Pvdd+*}), Eq.~(\ref{eq:FI-ir+*}) is close to zero everywhere except for these two peaks. The peaks are oriented in the $\phi_0$ direction. Furthermore, as it can be seen from the known behavior of the first Bessel function, the result remains finite even when the denominator in the second term inside the curly braces vanishes: the peaks described by this expression remain well-behaved everywhere in the region of weak interference, describing a light signal of finite intensity.

\section{Discussion and Conclusions}
\label{sec:disc}

We have studied the image formation process with the SGL and  analyzed the intensity distribution of the EM field received from a point source at the focal plane of an optical imaging telescope placed in the focal region of the SGL.

We first considered the SGL's region of strong interference. It is in this region, in the immediate vicinity of the SGL optical axis, where an image of a distant source is formed by the SGL. A commonly discussed mission concept \cite{Turyshev-etal:2018} envisions an optical telescope that will scan this region by moving laterally. Such a telescope must have sufficient angular resolution in order for it to benefit from a coronagraph, blocking out light from the Sun. The telescope will be able to capture an image of the Einstein ring that forms around the Sun from light received from a distant source. Investigating the propagation of the light field first through the SGL and then through the telescope optics, we were able to reconstruct the Einstein ring that appears in the focal plane of the telescope. We also verified that in the limit of vanishing solar gravitational field, the well-known Airy-pattern of the optical telescope emerges. Thus, our results extend all previously known results in the case of a monopole gravitational field.

As the telescope moves from the optical axis, it briefly passes through a region, characterized by diminishing partial arcs of the Einstein ring, where the integral expression that describes the amplitude of the image field can only be solved numerically. Outside this region, the partial arcs swiftly shrink to two spots, appearing on opposite sides of the Sun. This behavior was also successfully reconstructed analytically. Thus we were able to obtain analytic expressions for the image formed by an optical telescope in all the cases important for practical applications of the SGL for imaging distant sources.

In addition, we also considered a telescope situated at large distance away from the optical axis, both in the weak interference region (where two images of uneven brightness, on both sides of the Sun, are still present) and the geometric optics region (characterized by only one image, as light rays on the opposite side of the Sun are now blocked by the opaque solar disk.) We were able to provide a wave-optical treatment for gravitational microlensing phenomena, analytically reconstructing the asymmetric location of the two images and their uneven brightness.

As a result, we were able to describe the image formation process in all three regions of practical importance for imaging with the SGL, namely the strong and weak interference regions, and the region of geometric optics. Starting from the strong interference region, we see that in the case of very small deviations, characterized by $\alpha \rho_0\ll1$ and representing displacements in the range of $0\leq \rho_0\ll 1/\alpha\approx 2\sqrt{({\overline z}/650\,{\rm AU})}\,{\rm cm}$, an observer will see the intensity distribution in the form of the Einstein ring (\ref{eq:amp-w-f2dp}) that is formed in the focal plane of an optical telescope.  As the telescope moves further away from the optical axis,  an observer would first see the Einstein ring break into two arcs positioned on opposite sides of the Sun. And then, for large deviations, $\alpha \rho_0\gg1$, described as $\rho_0> 10/\alpha\approx 20\sqrt{({\overline z}/650\,{\rm AU})}\,{\rm cm}$, these arcs eventually morph into two peaks of identical brightness given by (\ref{eq:Pvdd+}).

As $\rho_0$ continues to increase, the two peaks move away from each other. The factor in front of the second term in (\ref{eq:FI-ir+}) leads to the eventual disappearance of the second term. Ultimately, however, this image is hidden by the Sun, and we enter the region of geometric optics characterized by (\ref{eq:FI-go}).

With the results derived in this paper, we are now at the position where we may begin to consider practical applications of the SGL. The next step is to  evaluate the signals that one may expect from various relevant sources. The same expressions may also be used to derive and study the instrument and mission requirements for a prospective mission to the focal region of the SGL. This work is underway  and results, when available, will be reported elsewhere.

\begin{acknowledgments}
This work in part was performed at the Jet Propulsion Laboratory, California Institute of Technology, under a contract with the National Aeronautics and Space Administration.
VTT acknowledges the generous support of Plamen Vasilev and other Patreon patrons.

\end{acknowledgments}


\end{document}